\documentclass[twocolumn]{aastex61}

\usepackage[utf8]{inputenc}
\usepackage{amsmath}
\usepackage{amsfonts}
\usepackage{amssymb}
\usepackage{graphicx}
\usepackage{natbib}
\usepackage{cleveref}
\usepackage{url}
\newcommand{\ly}{{Ly$\alpha$ forest}}
\newcommand{\lal}{{Ly$\alpha$}}

\defcitealias{Viel03}{V03}

\received{\today}
\revised{\today}
\accepted{\today}
\submitjournal{ApJ}

\shorttitle{DE and the Ly$\alpha$ Forest}
\shortauthors{Coughlin et al.}

\begin{document}

\correspondingauthor{Jared W.\ Coughlin}
\email{jcoughl3@illinois.edu}

\author[0000-0002-4373-4114]{Jared W.\ Coughlin}
\affiliation{NCSA, University of Illinois, Champaign, IL 61801}

\author{Grant J.\ Mathews}
\affiliation{Department of Physics, University of Notre Dame, Notre Dame, IN 46556}

\author{Lara Arielle Phillips}
\affiliation{Department of Physics, University of Notre Dame, Notre Dame, IN 46556}

\author{Ali P. Snedden}
\affiliation{The Research Institute at Nationwide Children's Hospital, Columbus, OH 43205}

\author{In-Saeng Suh}
\affiliation{Department of Physics, University of Notre Dame, Notre Dame, IN 46556}
\affiliation{Center for Research Computing, University of Notre Dame, Notre Dame, IN 46556}

\title{Probing Time-Dependent Dark Energy with the Flux Power Spectrum of the Lyman $\alpha$ Forest}

\begin{abstract}

We present new simulations of the flux power spectrum of the Lyman $\alpha$ forest as a means to investigate the effects of time-dependent dark energy on structure formation. We use a linearized parameterization of the time-dependence of the dark energy equation of state and sample the parameters ($w_0,w_a$) from the the extrema of the allowed observational values as determined by the Planck results. Each chosen ($w_0,w_a$) pair is then used in a high-resolution, large-scale cosmological simulation run with a modified version of the publicly available SPH code {\tt GADGET-2}. From each of these simulations we extract synthetic Lyman $\alpha$ forest spectra and calculate the flux power spectrum. We use the k-sample Anderson-Darling test to analyze the effects of dark energy on the Lyman $\alpha$ forest. We compare each dark energy power spectrum to that due to a cosmological constant power spectrum. We find, however, that there is only a marginal effect of the choice of allowed dark energy models on the flux power spectrum.

\end{abstract}

\keywords{cosmology: dark energy --- cosmology: theory --- methods: numerical}

\section{Introduction}
\label{sec:intro}

	Evidence of cosmic acceleration was first noted based upon Type Ia supernovae (SNIa) \citep{riess98,perlmutter99}, and has since been confirmed by multiple independent observational probes. These include measurements of baryon acoustic oscillations \citep[e.g.,][]{cole05,eisenstein05,WiggleZ11}, studies of the cosmic microwave background (CMB) anisotropies \citep[e.g.,][]{spergel03,komatsu11,planck13,planck15,Planck18}), the late-time integrated Sachs-Wolfe Effect (ISW) \citep[e.g.,][]{Dupe11,Ho08,Giannantonio08}, and weak gravitational lensing studies \citep[e.g.,][]{schrabback10,garcia-fernandez16,des_cosmic_shear, deJong13}).

	There are three possible explanations for cosmic acceleration. The first is that there is some exotic new mechanism known as dark energy driving the acceleration \citep{Ostriker95}. The second is that our understanding of gravity is not quite right and a modification to Einstein's general relativity is required \citep[e.g.,][]{Lue04,Carroll05}. The third is that we are living inside of a very under-dense region and there really is no cosmic acceleration at all; it simply appears as if there is \citep[e.g.,][]{Tomita00, Tomita01, Celerier00, Iguchi02, Enqvist08}. However, no convincing inhomogeneity model has been put forward to date \citep[e.g.,][]{AmendolaDE,Zhao11b}. Furthermore, modified gravity models are strongly restricted by local gravity constraints \citep[e.g.,][]{AmendolaDE}. As such, in this work we consider the dark energy paradigm as the most likely explanation. See, however, \citet{Joyce16} for an excellent discussion of dark energy as compared with modified gravity.

    Dark energy is commonly described by an equation of state (EOS) parameter $w(z) \equiv P_\mathrm{DE}/\rho_\mathrm{DE}$, where $P_\mathrm{DE}$ is the pressure and $\rho_\mathrm{DE}$ is the energy density for the dark energy. In this work we adopt natural units whereby $c = k_B = \hbar = 1$ unless otherwise noted. This EOS can be either constant or dynamical in time. Thus, when constraining models of dark energy, $w$ must be allowed to vary in time.

    Dynamical dark energy enters into the cosmic dynamics through the Friedmann Equation:
\begin{equation}
	\label{eq:friedmann-eq-dyn}
	\begin{split}
		H^2(a) &= H^2_0 \bigg [ \Omega_{r,0}a^{-4} + \Omega_{m,0}a^{-3} + \Omega_{K,0}a^{-2} + \\
		&\Omega_{\text{DE}, 0} \exp\left\{\int_a^1 \frac{3(1+w(a'))}{a'}da' \right\} \bigg ],
	\end{split}
\end{equation}
where $H$ is the Hubble parameter, $H_0$ is the present-day value of the Hubble parameter, $a$ is the cosmic scale factor, and $\Omega_{r,0}$, $\Omega_{m,0}$, $\Omega_{K,0}$, and $\Omega_{\text{DE},0}$ are the present-day values of the radiation, matter, curvature, and dark energy density parameters, respectively.

    Within the dark energy paradigm, there is a plethora of models that have been proposed as the physical mechanism responsible for cosmic acceleration \citep[e.g.,][]{AmendolaDE}, and they each posit a different form for $w(a)$. Two of the most popular are a simple cosmological constant (corresponding to $w(a) = -1$) \citep[e.g.,][]{Garnavich98} and a self-coupled, slowly evolving scalar field that usually manifests as either quintessence \citep{Caldwell98} or k-essence \citep{Armendariz-Picon00}. The empirical difference between the cosmological constant and all other dark energy models is that the energy density of the former remains constant as the Universe expands while the latter permits the energy density to vary in time. As such, observationally discriminating between various dark energy models reduces to determining how the dark energy changes in time. This can be done by studying the expansion history of the Universe and comparing observations with \Cref{eq:friedmann-eq-dyn}.

    Dark energy has only recently $(z \approx 0.3)$ come to dominate the dynamics of the Universe. This means that, for much of the history of the Universe, the effects of dark energy were sub-dominant to those of matter and radiation. This makes discrimination between dark energy models difficult since any purported time-dependent effects from dark energy will be small. In part because of this, current observational constraints on dark energy are still quite weak \citep[e.g.,][]{Planck15_DE}, with determinations of $w(z = 0)$ ranging from $w(z = 0) = -1.54^{+0.62}_{-0.50}$ to $w(z = 0) = -1.006^{+0.085}_{-0.091}$ \citep[see Table 5][]{planck15} depending on the combination of datasets being used as priors in their Bayesian analysis. This means that the exact form of $w(a)$ is unknown. Additionally, we currently lack the large number of measurements of $w(a)$ necessary for a nonparametric inference, and so we must use some parameterized form of $w(a)$ in order to compare with observations \citep[e.g.,][]{Upadhye05}. Here we employ the parameterization introduced in \cite{Linder03} and \cite{Chevallier01}:
\begin{equation}
	\label{eq:w-parameterization}
	w(a) = w_0 + w_a(1 - a) = w_0 + w_a \frac{z}{1+z},
\end{equation}
where $w_0$ is the present-day value of the equation of state $w(a=1)$ and $w_a$ is its derivative with respect to the scale factor $dw/da|_{a=1}$.

    Given the compelling evidence for cosmic acceleration, as well as it's unknown nature, it is imperative to explore every possible observational probe, as no one observational probe can adequately discriminate between models on its own \citep[e.g.,][]{Gerke02}. In this paper we present simulations that allow for this parameterization of dark energy and study whether or not dark energy imprints an observationally detectable signature on the flux power spectrum of the Lyman $\alpha$ (\lal{}) forest.

    The \ly{} is the dense collection of \ion{H}{1} \lal{} absorption lines in spectra of distant quasars (QSOs). Each of these lines occurs in the spectrum due to a particular absorber at a particular intervening redshift. The expansion of the Universe then causes each of these lines to redshift away from the \lal{} rest wavelength of $\approx 1216$\AA\ in accordance with the redshift of the absorber. See \citet{Rauch98} for an excellent review of the \ly{}.

    The motivation for using the \ly{} to study dark energy is as follows. The cosmic web \citep{bond96} is composed of three major types of structures: clusters, filaments, and voids. Since dark energy possesses a negative energy density, the effects of dark energy, with respect to the cosmic web, should be most apparent on the morphology of voids \citep[e.g.,][]{Park07, Bos12, Lee09, Biswas10, Lavaux10, Lavaux12, Shoji12}. The absorbers responsible for the \ly{} should reside primarily in the clusters and filaments \citep[e.g.,][]{Cen94, Bi95}. However, along a given line of sight (LOS), on average these absorbers will be separated by the voids. As such, the separation of these absorbers in redshift space should act as a tracer of the evolution of the voids \citep[e.g.,][hereafter referred to as \citetalias{Viel03}]{Viel03}. As such, the flux power spectrum, which acts as a proxy for the matter power spectrum \citep[e.g.,][]{McDonald00}, should contain information about how each dark energy model affects the cosmic acceleration. Specifically, dark energy affects the linear growth factor $D_1(z)$, which can be probed via the \ly{} \citep[e.g.,][]{Kujat02}. Additionally, studying the \ly{} is an independent and complimentary approach to searches for time-dependent dark energy based on the SNIa redshift-distance relation, the CMB, BAO, ISW, and gravitational lensing, and one that has received comparatively little attention in the literature \citep[see, however,][]{Viel03,Greig_thesis,Kujat02}.

    This effect of dark energy has been explored previously in \citetalias{Viel03}, however they did not consider fully dynamical dark energy, instead focusing on various values of constant $w(a)$. Additionally, \citetalias{Viel03} used a semi-analytic treatment in their study of the \ly{}. Here we expand upon their work in two ways. First, we make use of high-resolution N-body simulations, from which we extract synthetic \lal{} spectra, and second we use fully dynamical models of dark energy that probe the currently allowed parameter space for $(w_0, w_a)$.

	This paper is organized in the following manner: Section~\ref{sec:sims} presents the details of our simulations, Section~\ref{sec:num_methods} gives a description of the numerical procedures adopted in order to generate our synthetic spectra, results are presented in Section~\ref{sec:results}, and conclusions are summarized in Section~\ref{sec:discussion}

\section{Simulations}
\label{sec:sims}
	Our simulations were run with a modified version of the publicly available smoothed particle hydrodynamics (SPH) code {\small GADGET-2} \citep{springel05c,Dolag04,snedden16}.

	Simulating the \ly{} requires very high resolution. It has been suggested \citep{McDonald03} that a resolution of $\approx 40\;h^{-1}\;\text{kpc}$ in a box of size $L \approx 40\;h^{-1}\;\text{Mpc}$ is needed in order to adequately resolve the structure of the \ly{} and achieve convergence for the power spectrum. With these requirements in mind, we simulated $1024^3$ dark matter particles in a box of length $40\;h^{-1}$ comoving Mpc. This gives a particle mass of $5.21 \times 10^6 M_\sun\; h^{-1}$.

	Due to the high resolution requirements of our simulations, we only evolve a distribution of dark matter particles out of consideration for the total run-time. This is justified because the baryon distribution largely follows that of the dark matter on large scales \citep[e.g.,][]{Meiksin01} in the low-density, mildly non-linear environments typically responsible for the \ly{}. Additionally, the effects of non-linear baryonic physics, such as galactic winds, have been shown to be small at large scales \citep[e.g.,][]{Bertone06} where the effects of dark energy should be most prominent. See, however, Section~\ref{sec:discussion}.

    Our cosmological parameters are those given in the \cite{planck15} analysis and summarized in \Cref{table:sim-params} along with the other simulation parameters of note. We list the number of SPH neighbors in \Cref{table:sim-params}. The neighbors are used in the post-processing when calculating the densities. This is necessary for extracting synthetic \ly{} spectra, as described in Section~\ref{sec:dens-calc}.

\begin{figure*}
	\centering
    \includegraphics[width=\textwidth]{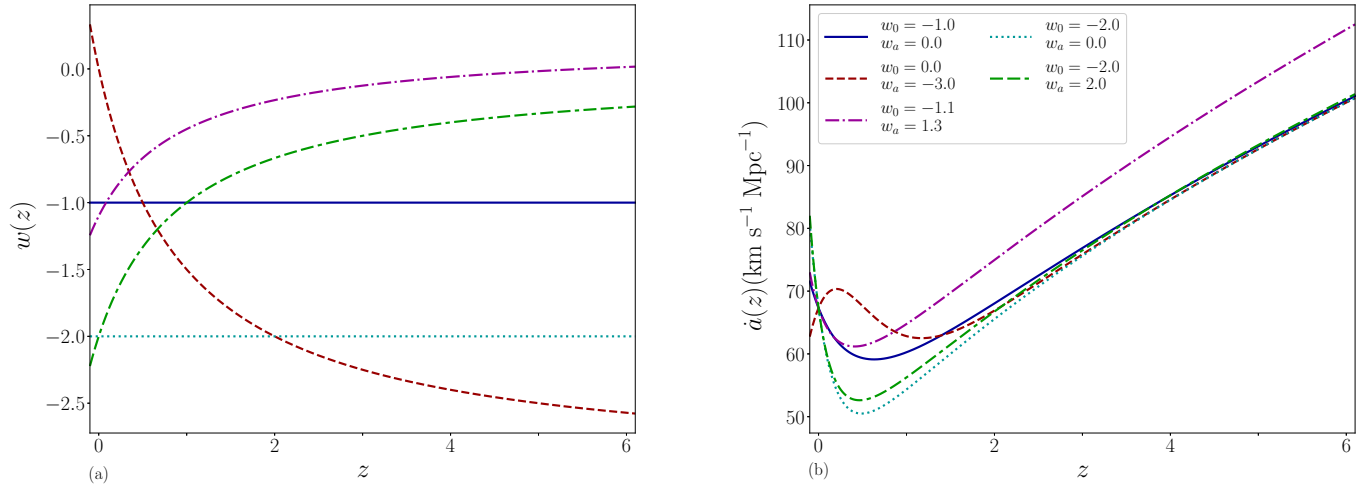}
    \caption{(a): The evolution of the dark energy EOS $w(z)$ for each of the dark energy models that we consider in this paper. The blue line corresponds to the cosmological constant ($w_0 = -1, w_a = 0$). The other models were chosen to be close the edges of the 95\% confidence $w_0-w_a$ parameter space as determined by \citet{Planck15_DE}, save for model DE2-40-1024 ($w_0 = -1.1,\; w_a = 1.3$), which was deliberately chosen to be outside of the allowed range. See text for details. We see that the EOS for different dark energy models can vary considerably in their behavior, thereby affecting the expansion history of the Universe in unique ways via \Cref{eq:friedmann-eq-dyn}. (b): The time derivative of the cosmic scale factor as a function of redshift for each of the dark energy models considered in this paper. This panel highlights the differences in the expansion history due to dark energy. The color coding and legend are the same for both panels.}
    \label{fig:w-plot}
\end{figure*}

	We ran five simulations\footnote{Any and all of our simulation data are available upon request.}: one with a cosmological constant and four with dynamical dark energy where $w(a)$ was given by \Cref{eq:w-parameterization}. The dynamical models, shown in \Cref{fig:w-plot} along with the cosmological constant, were chosen such that their parameters were at the edges of the allowed $95$\% confidence range for the $(w_0,w_a)$ parameter space given in \cite{Planck15_DE}. We chose to be at the fringes of the allowed parameter space as these models should produce flux power spectra with the largest differences between them. The only exception to this is model DE2-40-1024, which was deliberately chosen from a region of the $(w_0, w_a)$ parameter space that is outside the $95$\%5 confidence level bounds given by the Planck marginalized posteriors. This was done for two reasons: first, we wanted to determine if very extreme values of the dark energy parameters were capable of producing a distinct signature in the flux power spectrum of the \ly{}, and second, we wish to determine whether or not the flux power spectrum provides constraints on the dark energy parameter space that are in accord with the results determined from other observational probes. See Section~\ref{sec:discussion}. The values of $w_0$ and $w_a$ that we considered are summarized in \Cref{table:dark-params}.

	Each of our simulations began from the same initial conditions and was evolved from $z=49$ to $z=2.2$. Our initial conditions were generated using the publicly available second-order Lagrangian perturbation theory code {\tt 2LPTIC} \citep{scoccimarro12}. We generated snapshots of each simulation for quasars at $z=4.2,3.8,3.0,2.7,$ and $z=2.2$. We chose these particular redshift values because they correspond to the redshifts at which there are determinations of the \ly{} flux power spectrum from multiple observations, as described in Section~\ref{sec:results}. Our simulations required on the order of $\approx 10$ days on $72$ processors to run. Following this, the post-processing required on the order of one week per snapshot, with the majority of the time (about four or five days) devoted to halo-finding (see Section~\ref{sec:temp-calc}).
\begin{deluxetable}{lr}
    \tablewidth{0pt}
    \tabletypesize{\scriptsize}
    \tablecolumns{2}
    \tablecaption{Simulation Parameters\label{table:sim-params}}
    \tablehead{\colhead{Simulation Parameter} & \colhead{Value}}
    \startdata
        Number of DM Particles & $1024^3$ \\
        Number of SPH Neighbors (tolerance) & $48$($\pm 3$) \\
        Softening Length ($\mathrm{kpc\,h}^{-1}$ comoving) & $1.56$ \\
        Box Size ($\mathrm{Mpc\,h}^{-1}$ comoving) & $40.0$ \\
        Particle Mass & $5.21 \times 10^{6}\mathrm{M}_\sun h^{-1}$ \\
        $z_\text{start}$ & $49.0$ \\
        $z_\text{end}$ & $2.2$ \\
        $\Omega_{m,0}$ & $0.315$ \\
        $\Omega_{b,0}$ & $0.0456$ \\
        $\Omega_{\text{DE},0}$ & $0.685$ \\
        $H_0$ ($\mathrm{km\,s}^{-1}\mathrm{Mpc}^{-1}$) & $67.31$ \\
    \enddata
\end{deluxetable}

\begin{deluxetable}{lrr}
    \tablewidth{0pc}
    \tabcolsep=25pt
    \tablecolumns{3}
    \tablecaption{Simulation Dark Energy Parameters\label{table:dark-params}}
    \tablehead{\colhead{Simulation Name} & \colhead{$w_0$} & \colhead{$w_a$}}
    \startdata
        L-40-1024 & $-1.0$ & $0.0$ \\
        DE1-40-1024 & $0.0$ & $-3.0$ \\
        DE2-40-1024 & $-1.1$ & $1.3$ \\
        DE3-40-1024 & $-2.0$ & $0.0$ \\
        DE4-40-1024 & $-2.0$ & $2.0$ \\
    \enddata
\end{deluxetable}

\section{Numerical Methods}
\label{sec:num_methods}
	Here we present the details of our modifications to {\small GADGET-2}, our spectral extraction code, and our power spectrum calculation.\footnote{All of our code can be found at \url{https://bitbucket.org/polaris42/}.}

\subsection{Implementation of Dynamical Dark Energy}
\label{sec:dyn_de_implementation}
	 The publicly available version of {\small GADGET-2} assumes that dark energy arises from a cosmological constant. Modifying {\small GADGET-2} to include the effects of dynamical dark energy is relatively straight forward. This is because using the parameterization given in \Cref{eq:w-parameterization} in \Cref{eq:friedmann-eq-dyn} gives rise to an analytic expression:
\begin{equation}
	\label{eq:friedmann-eq-w-de-parameterization}
	\begin{split}
		H^2(a) &= H^2_0 \bigg [ \Omega_{r,0}a^{-4} + \Omega_{m,0}a^{-3} + \Omega_{K,0}a^{-2} + \\
		&\Omega_{\text{DE}, 0} \left\{a^{-3 (1+w_0+w_a)}e^{-3w_a(1-a)}\right\} \bigg ].
	\end{split}
\end{equation}
We then call \Cref{eq:friedmann-eq-w-de-parameterization} whenever dynamical dark energy is used. The parameters $w_0$ and $w_a$ are read in from the {\small GADGET-2} parameter file. For reasons of speed, this was done using a look-up table that was generated at the beginning of the run.

\subsection{Generation of Synthetic Spectra}
\label{sec:syn-spec}
	Calculating a synthetic spectrum requires the densities, temperatures, and \ion{H}{1} fractions for each of the simulation particles. Since these are not properties of the dark matter particles in our version of {\small GADGET-2}, we calculate these quantities in post-processing.

\subsubsection{Density Calculation}
\label{sec:dens-calc}
	The densities were determined using the {\small GADGET-2} density calculation \citep{springel05c} adapted to work for dark matter particles. The density $\rho_i$ of particle $i$ is calculated using:
\begin{equation}
	\label{eq:sph-dens}
	\rho_i = \sum_{j=1}^N m_j W(|\mathbf{r}_{ij}|, h_i),
\end{equation}
where $\mathbf{r}_{ij} = \mathbf{r}_i - \mathbf{r}_j$ are the separations of particles $i$ and $j$, $h_i$ is the smoothing length of particle $i$, and $W(|\mathbf{r}_{ij}|, h_i)$ is the smoothing kernel. The smoothing lengths are calculated in {\small GADGET-2} by ensuring that the total amount of mass within the smoothing sphere remains approximately constant, i.e.,
\begin{equation}
	\label{eq:smoothing-lengths}
	\frac{4\pi h_i^3 \rho_i}{3} = N_\text{sph} \bar{m},
\end{equation}
where $N_\text{sph}$ is the user-defined typical number of neighbors to be enclosed by the particle's smoothing sphere and $\bar{m}$ is the average particle mass. Here we use $48 \pm 3$ neighbors \citep{dehnen12}. Additionally, we use the same polynomial smoothing kernel as in \citet{springel05c}
\begin{equation}
	\label{eq:smoothing_kernel}
	W(r,h) = \frac{8}{\pi h^3} \begin{cases} 1 - 6\left (\frac{r}{h} \right)^2 + 6\left (\frac{r}{h}\right )^3 & 0 \leq \frac{r}{h} \leq \frac{1}{2},\\
		2 \left (1 - \frac{r}{h} \right )^3 & \frac{1}{2} < \frac{r}{h} \leq 1,\\
		0 & \frac{r}{h} > 1.
		\end{cases}
\end{equation}
Using the {\small GADGET-2} tree, we determined a list of nearest neighbors for each dark matter particle. This list then allows for \Cref{eq:sph-dens,eq:smoothing-lengths} to be solved.

\subsubsection{Temperature Calculation}
\label{sec:temp-calc}
	We consider two different environments for our simulation particles: halo and field (where field particles are those not identified as belonging to halos and correspond to the inter-galactic medium (IGM)). We make this distinction because the physical conditions of the two environments differ significantly, so we calculate the temperature of each particle based upon which environment it resides in \citep[e.g.,][]{Bertone_thesis,Popping09,Duffy12}.

    Particles in the field are generally unshocked. They undergo adiabatic cooling and are photoionized by the UV background \citep{Bertone_thesis}. \cite{Hui97} showed that there exists a power-law relation between the temperature and the density of the gas in the IGM:
\begin{equation}
	\label{eq:igm-eos}
	T = T_0 (1 + \delta)^{\gamma - 1},
\end{equation}
where $T_0$ \citep{theuns98} is the temperature of the IGM at a given redshift and mean density for that redshift. The slope of the power law is given by $\gamma - 1$, and we use $\gamma - 1 \approx 1/1.7$ \citep{Irsic14}. The over-density $\delta = (\rho - \bar{\rho})/\bar{\rho}$, where $\bar{\rho}$ is the mean density as a function of redshift, is given by
\begin{equation}
	\label{eq:mean-dens}
	\bar{\rho}(z) = \Omega_m(z) \rho_c(z) = \frac{\Omega_{m,0} (1+z)^3\rho_c(z)}{E(z)^2},
\end{equation}
where $\rho_c$ is the critical density and $E(z) = H(z)/H_0$ is the expansion factor. \Cref{eq:igm-eos} is the result of considering low-density regions in photoionization equilibrium and adiabatic cooling. The average temperature of the IGM at redshift $z$ is given by

\begin{equation}
	\label{eq:T0}
	\begin{split}
		T_0(z) &= \Bigg \{\left[\frac{\Omega_b h t'_H L'_\epsilon (1 + z)^\frac{3}{2}}{t'_\text{heat}(2+\omega)}\right] \bigg/\\
		&\left [1 - \frac{L'_{cc}t'_H(1+z)^\frac{5}{2}}{ht'_\text{heat}}\right ] \Bigg \}^{\gamma - 1},
	\end{split}
\end{equation}

	Where $\Omega_b$ is the density parameter for baryons, $h$ is the little h parameter \citep{croton13} used in the definition of the Hubble parameter $H_0 = 100\;h\;\text{km s}^{-1}\;\text{Mpc}^{-1}$, and
\begin{align}
	 &L'_\epsilon =1.7\times 10^{-20}\text{erg s}^{-1}\text{cm}^3\text{K}^{0.7}, \\
	 &L'_cc = -7.31\times 10^{-30}\text{erg s}^{-1}\text{cm}^3\text{K}^{-1.0}, \\
	 &t'_\text{heat} = 5.41\times 10^{-11}\text{erg cm}^{3}\text{K}^{-1.0}, \\
	 &t'_H = 2.06\times 10^{17}\text{s},
\end{align}
with $\omega = 1.5$ \citep{Bertone_thesis}.

    Halo finding was done using the Amiga Halo Finder ({\sc Ahf}) \citep{knollmann09,Gill04}. {\sc Ahf} is a parallel and publicly available halo finding code that identifies halos through a hierarchical grid generated through adaptive mesh refinement. This procedure has several advantages, chief of which is the ability to naturally identify sub-structure within each halo via the more refined grid levels. The ability to use substructure is important as it allows for the temperature calculation to be refined through the use of the properties of the sub-halo as opposed to being restricted to the properties of the host. Additionally, {\sc Ahf} has the ability to incorporate non-standard dark energy models into its halo calculation out of the box. This makes it ideally suited for our purposes.

	Within halos, the gas is shock heated due to the non-linear processes of structure formation. Additionally, self-shielding occurs. This affects how the particles interact with the UV background. Since halo particles do not follow \Cref{eq:igm-eos}, their temperatures are calculated from the virial properties of the halo in which they reside. We require a minimum of $1000$ particles in a halo. This implies a minimum halo mass of $\approx 10^9M_\sun$, which is reasonably consistent with observation (e.g. \cite{Bordoloi14,Brook14,Gerhard92}).

	The virial properties of the halos are given by \citep{Mo02}:
\begin{align}
	&R_\text{vir} = \left [ \frac{GM_\text{vir}}{100 H^2 (z)} \right ] ^\frac{1}{3}, \label{eq:virial-r}\\
	&V_c = \left ( \frac{GM_\text{vir}}{R_\text{vir}} \right ) ^\frac{1}{2}, \label{virial-v}\\
	&T_\text{vir} = \frac{\mu m_p V_c^2}{2k_B}, \label{virial-T}
\end{align}
where $R_\text{vir}$ is the virial radius (the radius at which the included mass gives rise to a mean over-density of two-hundred times the critical density), $V_\text{vir}$ is the circular velocity, $M_\text{vir}$ is the virial mass, $T_\text{vir}$ is the virial temperature, $\mu$ is the mean molecular weight, and $m_p$ is the mass of the proton. We use $\mu = 0.588$. This value of the mean molecular weight assumes a primordial composition of $X = 0.76$ and $Y=0.24$, where $X$ is the hydrogen mass fraction and $Y$ is the helium mass fraction.

\begin{figure}
	\centering
    \includegraphics[width=\columnwidth]{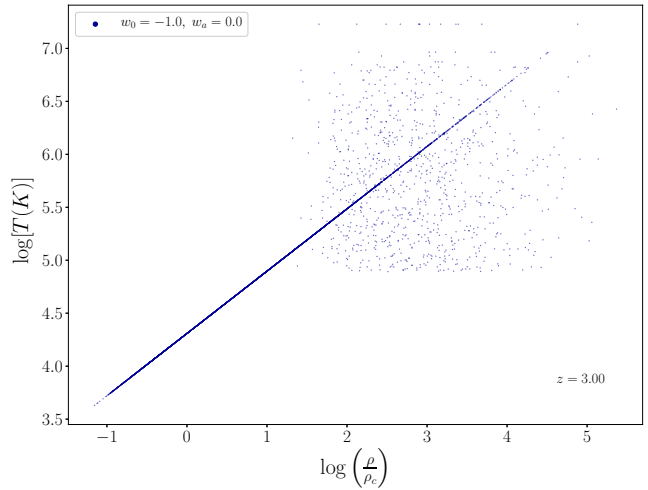}
	\caption{The $\log (T)$ vs. the $\log \left (\rho/\rho_c \right)$ for the L-40-1024 simulation at $z=3$. The straight line is due to the field particles obeying \Cref{eq:igm-eos} and the halo particles fall into the space given by the block-like region. For visual clarity, we only plot $5\times 10^{-4}$ percent of the total number of particles in the simulation.}
	\label{fig:T-vs-rho}
\end{figure}

	\Cref{fig:T-vs-rho} shows the temperature-density relation for a subset of the total number of simulation particles. The figure shows two distinct groups: the straight line and the block-like region. These two groups are artifacts of our bimodal temperature calculation scheme. The straight line represents the field particles whose temperatures are calculated from \Cref{eq:igm-eos}, and the block-like region is due to the halo particles. The temperatures of the halo particles tend to fall in horizontal bands across the region. This is because every particle in the same halo has its temperature calculated from the same virial properties. Hence, they have nearly the same temperature. See \cite{Bertone06}.

\subsubsection{HI Neutral Fractions}
\label{sec:XHI}
	From our computed temperatures and densities we can deduce the fraction of each particle's mass that is in the form of neutral hydrogen. In order to perform this calculation, we evolved the hydrogen mass as a difference between recombination and ionization rates, i.e.
\begin{equation}
	\label{eq:H-evolution}
	\frac{dX_\text{HI}}{dt} = \alpha_\text{HII}n_\text{e}X_\text{HII} - X_\text{HI} (\Gamma_{\gamma,\text{HI}} + \Gamma_{e,\text{HI}}n_\text{e}),
\end{equation}
where $X_\text{HI} = n_\text{HI}/n_\text{H}$ is the neutral hydrogen fraction, $n_i$ is the number density of species $i$, $n_\text{e} = \text{\emph{e}}n_\text{H}$ is the electron number density, \emph{e} is the electron fraction, $X_\text{HII}$ is the fraction of ionized hydrogen, $\alpha_\text{HII}$ is the recombination rate of ionized hydrogen, $\Gamma_{\gamma,\text{HI}}$ is the photoionzation rate of neutral hydrogen, and $\Gamma_{e\text{HI}}$ is the collisional ionization rate of neutral hydrogen. The electron fraction is given by \citep{Bertone_thesis}:
\begin{equation}
e = \frac{2-Y}{2(1-Y)}.
\end{equation}

	To solve \Cref{eq:H-evolution}, we assume ionization equilibrium, for which $dX_\text{HI}/dt = 0$. Furthermore, we assume that the gas is initially highly ionized, i.e. $X_\text{HII} \approx 1$. These are reasonable assumptions when modeling the IGM \citep[e.g.,][]{schaye08}. With these assumptions the neutral hydrogen fraction is given by:
\begin{equation}
	\label{eq:neutral-H}
	X_\text{HI} = \frac{\alpha_\text{HII}n_\text{e}}{\Gamma_{\gamma,\text{HI}} + \Gamma_{e,\text{HI}}n_\text{e}},
\end{equation}
and the recombination and ionization rates are given by \citep{theuns98}:
\begin{align}
	&\alpha_\text{HII} = 6.3\times 10^{-11} \frac{T^{-0.5}T_3^{-0.2}}{1+T_6^{0.7}},\\
	&\Gamma_{\gamma\text{HI}}(z) = \int_{\nu_i}^\infty \frac{4\pi J(\nu, z)\sigma_{\text{HI}}(\nu)}{h\nu}d\nu,\\
	&\Gamma_{e\text{HI}} = 1.17\times 10^{-10}T^{0.5}\exp\left[\frac{-157809.1}{T}\right]\frac{1}{1+T_5^{0.5}},
\end{align}
where $T_n \equiv T/10^nK$, $J(\nu,z)$ is the photoionizing UV background, $\sigma_\text{HI}(\nu)$ is the photoionization cross section for neutral hydrogen, and $h$ is Planck's constant reintroduced here for clarity. In this work we adopt the UV background of Haardt and Madau 2005 \citep{haardt96}.

	In reality, the actual form of the UV background is still highly uncertain \citep[e.g.,][]{McQuinn16}. The exact form of $J$ is believed to be due to contributions from both quasars and star forming galaxies \citep[e.g.,][]{Haardt12}. While there are constraints on the quasar contribution, the contribution from star forming galaxies is far more uncertain, due mostly to the fact that the fraction of UV photons that can escape from star forming galaxies is unknown \citep[e.g.,][and references therein]{Rivera-Thorsen17, Vanzella18}. Additionally, \citet{Madau99} have shown that the known population of quasars and galaxies cannot provide enough UV photons to produce the ionization state of the present-day IGM. This large uncertainty surrounding the UV background affects the numerical procedure for spectral extraction via the normalization of each spectrum, as discussed in section~\ref{sec:spec-extract}.

\subsubsection{Spectral Extraction}
\label{sec:spec-extract}
	For simplicity, each LOS is taken parallel to the x-axis of the simulation volume. However, the starting point on the face of the cube is chosen randomly. We call the starting point of the LOS point A, and it corresponds to the location of the quasar responsible for the synthetic spectrum we are generating. We call the end of the LOS point B, which corresponds to the location of the observer who would see the synthetic spectrum we are generating. These two points are different for each spectrum we generate. Each LOS is then broken up into $N$ segments, which we call pixels. We use $N=3000$, as we found that this number gives a good balance between spectral resolution and run-time.

    In order to generate a spectrum along a LOS, the optical depth $\tau$ of each pixel must first be calculated. This involves determining the temperatures, densities, and velocities of neutral hydrogen within each pixel. Calculating these pixel quantities requires calculating the contribution of each particle to each pixel. These contributions are given as \citep{theuns98, Bertone06}:
\begin{align}
	\label{eq:pix-rho-HI}
	&\rho_{X,j} = \sum_i X_i m_i W(|\mathbf{r}_{ij}|, h_i),\\
	\label{eq:pix-rho-T}
	&(\rho T)_{X,j} = \sum_i X_i m_i W(|\mathbf{r}_{ij}|, h_i) T_i,\\
	\label{eq:pix-rho-v}
	&(\rho v)_{X,j} = \sum_i X_i m_i W(|\mathbf{r}_{ij}|, h_i) v_{\text{tot,LOS},i},
\end{align}
where, for each pixel $j$, $\rho_{X,j}$ is the density of \ion{H}{1}, $(\rho T)_{X,j}$ is the density-weighted temperature, and $(\rho v)_{X,j}$ is the density-weighted velocity. $X$ refers to the mass fraction of the ion being considered (in our case, we are only considering \ion{H}{1}, so $X$ is the fraction of neutral hydrogen). The sums are over all of the particles $i$ that overlap pixel $j$. This makes $T_i$ the temperature of particle $i$, $v_{\text{tot,LOS},i}$ the total velocity (Hubble velocity plus peculiar velocity) of particle $i$ directed along the LOS, and we take the location of pixel $j$ to be the center of the pixel.

	Once the pixel quantities described in \Cref{eq:pix-rho-HI,eq:pix-rho-T,eq:pix-rho-v} have been found, they can be used to evaluate the optical depth. The absorption that occurs in a given pixel $k$ is due to both the gas in that pixel plus the gas in other pixels that has been shifted due to peculiar motion:
\begin{equation}
\label{eq:optical-depth}
\tau_k = \sum_j \frac{\sigma_\alpha c n_{\text{HI},j}a \delta}{\sqrt{\pi}b_{\text{HI},j}} \exp\left[ -\left (\frac{v_{H,k} - v_{\text{HI},j}}{b_{\text{HI},j}} \right )^2 \right ],
\end{equation}
where the speed of light $c$ has been reintroduced for clarity. The sum is over all of the pixels, and $n_{\text{HI},j}$ is the neutral hydrogen number density of pixel $j$. This is found by dividing \Cref{eq:pix-rho-HI} by the mass of hydrogen. The quantity $\sigma_\alpha = 4.45\times 10^{-18}\text{cm}^2$ is the cross-section for the \lal{} transition in neutral hydrogen and $v_{\text{HI},j}$ is the velocity of the neutral gas. This is found by dividing \Cref{eq:pix-rho-v} by \Cref{eq:pix-rho-HI}. The quantity $v_{H,k}$ is the Hubble velocity of pixel $k$, given by

\begin{equation}
	\label{eq:pix-hub-vel}
	v_{H,k} = H(z_k)d_k,
\end{equation}
where $d_k$ is the distance between pixel $k$ and point B, and $H(z_k)$ is the Hubble parameter at the redshift of pixel $k$. We find the redshift of each pixel from the method presented in \citet{trident}. Briefly, this method requires that we first find the redshift extent of the simulation volume, where $z_A$ is the redshift of the quasar (which is taken to be the redshift of the snapshot), by solving

\begin{equation}
	\label{eq:box-redshift-range}
    \frac{c}{H_0} \int_{z_B}^{z_A} \frac{dz'}{E(z')} - L = 0
\end{equation}
for $z_B$, which is the redshift of the observer. We solve \Cref{eq:box-redshift-range} using Newton's method. Once the velocity extent of the box is known, this allows us to assign a redshift to each pixel along the LOS according to

\begin{equation}
	\label{eq:pix-z}
	z_k = z_A - \sum_{i=0}^{k-1} dz_i,
\end{equation}
where

\begin{equation}
	\label{eq:dz-pix}
    dz_i = -\frac{\Delta_i}{l}(z_B - z_A),
\end{equation}
and $\Delta_i = \frac{l}{N}$ is the width of pixel $i$ with $l$ the length of the LOS.

	The Doppler parameter $b_{\text{HI},j}$ is given by:
\begin{equation}
\label{eq:doppler-param}
b_{\text{HI},j} = \sqrt{\frac{2 k_B T_{\text{HI},j}}{m_H}}.
\end{equation}
Here, $T_{\text{HI},j}$ is the temperature of the neutral hydrogen. This is found by dividing \Cref{eq:pix-rho-T} by \Cref{eq:pix-rho-HI}.

	Once the optical depth has been calculated for every pixel along the LOS, we then convert this to flux via:
\begin{equation}
	\label{eq:flux}
	F_i = e^{-\tau_i},
\end{equation}
where $F_i$ is the flux of pixel $i$ and $\tau_i$ is the optical depth of pixel $i$.

	We then get the observed wavelength of each pixel $\lambda_i$ using:
\begin{equation}
	\label{eq:z-lambda}
	\lambda_i = \lambda_0 (1+z_{\mathrm{eff},i}),
\end{equation}
where $\lambda_0$ is the rest wavelength of the Lyman $\alpha$ transition corresponding to $1215.6$\r{A} and $1 + z_{\mathrm{eff},i} = (1 + z_i) (1 + z_{\mathrm{Dop},i})$ \citep[e.g.,][]{trident}, where $z_{\mathrm{eff},i}$ is the effective redshift of the pixel $i$ that combines both the cosmological and Doppler redshifts of the pixel. The cosmological redshift of each pixel is given by \Cref{eq:pix-z} and the Doppler redshift of each pixel is given by \citep[e.g.,][]{trident}

\begin{equation}
	1 + z_{\mathrm{Dop}, i} = \frac{1 + \frac{v_{\mathrm{los}, i}}{c}}{\sqrt{1 - \left(\frac{v_i}{c}\right)^2}},
\end{equation}
where $v_{\mathrm{los}, i}$ is the peculiar velocity of the gas in pixel $i$ directed along the LOS and $v_i$ is the magnitude of the peculiar velocity of the gas in pixel $i$.

	Each spectrum is then normalized by requiring the mean simulated flux to match the mean observed flux at the redshift in question. Here, we use the mean observed optical depth as given by \cite{Kim02}:
\begin{equation}
	\label{eq:mean-obs-tau}
	\bar{\tau}_\text{HI}(z) = 0.0032(1+z)^{3.37}.
\end{equation}
	The normalization is computed using an iterative process that continually rescales the average simulated flux and then compares it to the average observed flux calculated using \Cref{eq:mean-obs-tau} until the two converge to within 1\% of one another. This normalization procedure can be thought of as a rescaling of the UV background. This is needed due to the uncertainties surrounding the local UV flux (see Section~\ref{sec:XHI}).

\subsection{The Flux Power Spectrum}
\label{sec:flux-powspec}
	The absorption in the \ly{} serves as a means to map out the large-scale structure between the observer and the distant quasar source. Thus, the flux power spectrum serves as a proxy for the power spectrum of the underlying matter field that gives rise to the absorption in the \ly{}. Since the matter power spectrum is a measure of the density amplitudes as a function of scale, and these amplitudes depend upon the expansion history of the Universe, one can, in principle, use the flux power spectrum to discriminate between dark energy models.

	Following \cite{Hui01}, we do not analyze the flux directly but instead consider the quantity:
\begin{equation}
	\label{eq:pow-spec-quantity}
	F_p = \frac{e^{-\tau}}{\langle e^{-\tau}\rangle} - 1.
\end{equation}
$F_p$ is used in place of $F$ because $F$ is sensitive to changes in the mean flux $\langle e^{-\tau}\rangle$ \citep{Hui01}. We then take the Fourier transform of $F_p$ using the publicly available package {\tt FFTW3}\footnote{\url{http://www.fftw.org/}} to calculate the power spectrum, which we denote by $F_{p,k}$:
\begin{equation}
	\label{eq:powspec}
	P_{F_p}(k) = \mathcal{N}|F_{p,k}|^2.
\end{equation}
The normalization $\mathcal{N}$ of the spectrum is found by dividing out the total counts that occur within each bin of frequency $k$ and multiplying by the length of the \ly{} spectrum in velocity space. The frequencies are found via:
\begin{equation}
	\label{eq:k}
	k_i = \frac{2\pi i}{T},
\end{equation}
where $i$ indicates the bin index and $T$ is the period. For discrete Fourier Transforms (DFTs), the signal is assumed to be periodic over the range in which there is data, so the period is simply the length of the spectrum in velocity space. We bin our power spectra in the same manner as \citet{McDonald00}.

\section{Results}
\label{sec:results}
\subsection{Synthetic Spectra}
\label{sec:spec-results}

	Using the procedure described in Section~\ref{sec:spec-extract}, we extract $1152$ synthetic spectra for each model at each redshift.

\begin{figure*}
	\centering
    \includegraphics[width=\textwidth]{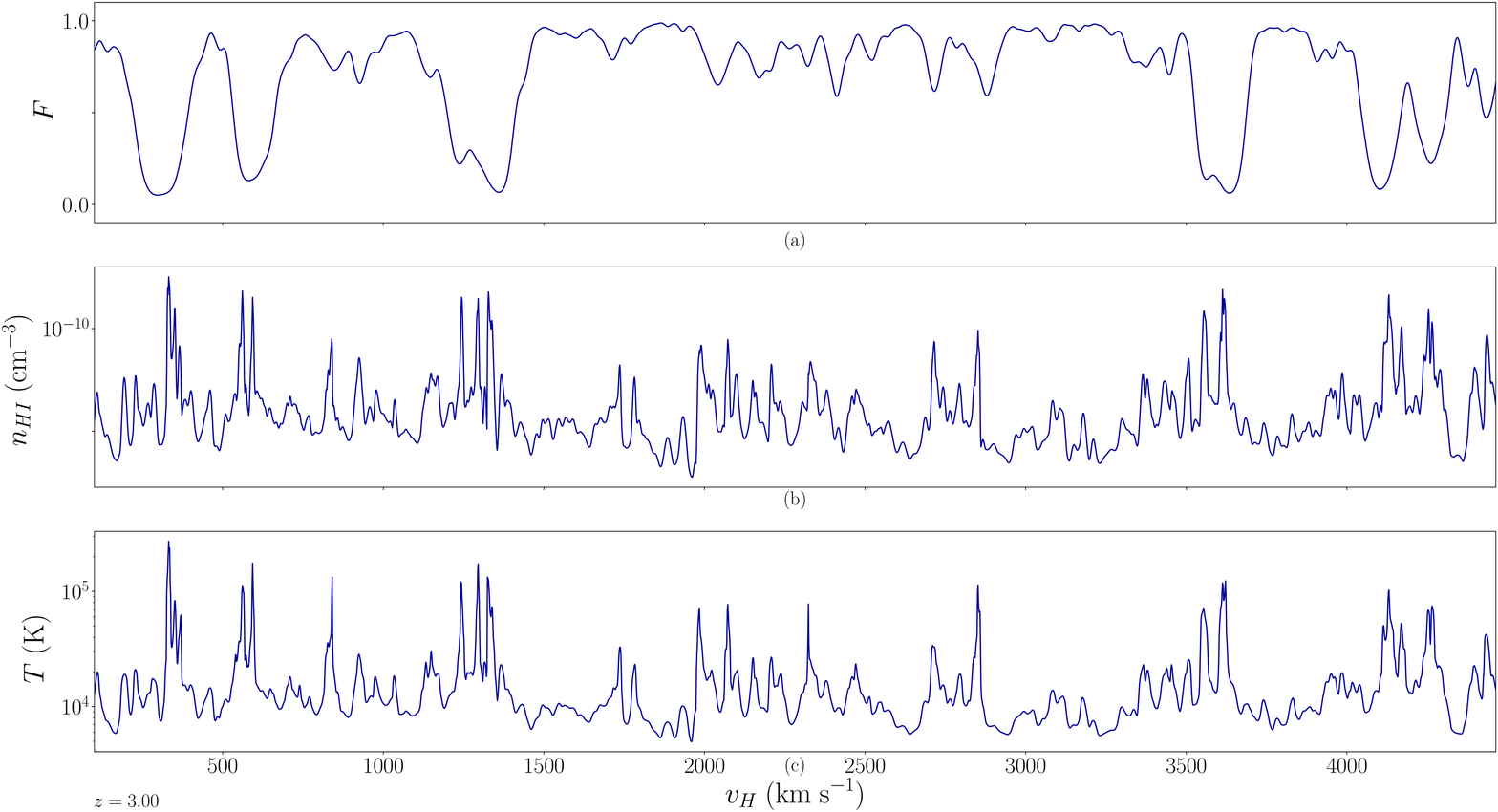}
    \caption{Example of a synthetic spectrum extracted from the center of our L-40-1024 simulation at $z = 3.00$. Panel (a) shows the flux $F = \exp^{-\tau}$ along the LOS, panel (b) shows the number density of neutral hydrogen along the LOS, and panel (c) shows the temperature. The x-axis is the same for each of the three panels.}
    \label{fig:example-spectrum}
\end{figure*}

	\Cref{fig:example-spectrum} shows an example of a spectrum that passes through the center of our L-1024-40 simulation volume. In addition to the flux we also plot the number density of neutral hydrogen and temperature of the pixels along the LOS. We see that this figure serves as a consistency check in that when there is a trough in the transmitted flux, there is a peak in the optical depth, as expected. Additionally, the temperatures of the pixels fall mostly near $\approx 10^4\mathrm{K}$, which is appropriate for the neutral hydrogen giving rise to the \ly{} \citep[e.g.,][]{Becker11}.

\begin{figure*}
	\centering
    \includegraphics[width=\textwidth]{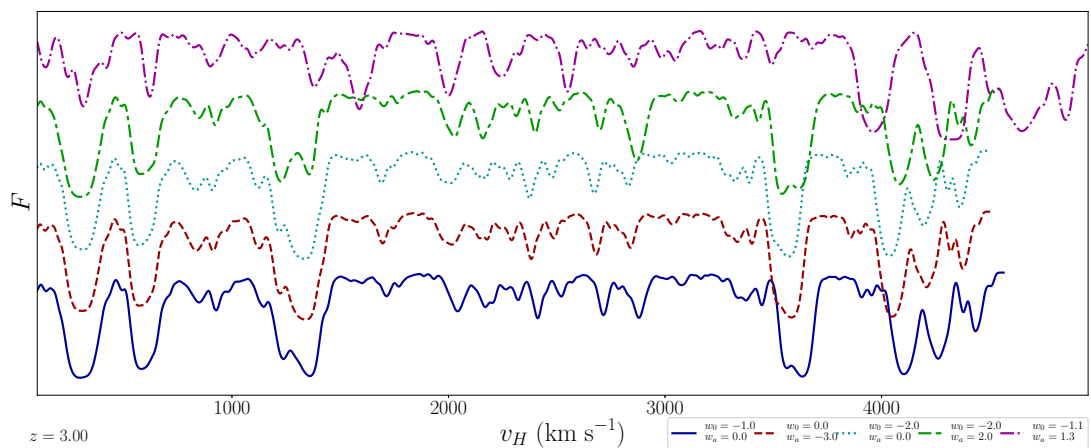}
    \caption{This figure shows the synthetic spectrum along the same LOS for each of our dark energy models at $z = 3.00$. This figure highlights the fact that the spectrum for each of our dark energy models, save for DE2-40-1024 (magenta), are all very similar to one another. Each spectrum is offset vertically from the others for reasons of visual clarity.}
    \label{fig:example-spectrum-all-flux}
\end{figure*}

    \Cref{fig:example-spectrum-all-flux} illustrates a synthetic \ly{} spectrum through the center of the simulation volume for each of our dark energy models. These are offset from one another for visual clarity. All of the spectra, except for the DE2-40-1024 model, are quite similar to one another.

\subsection{Power Spectra}
\label{sec:ps-results}
	In order to calculate a power spectrum at each redshift, we bootstrapped a sample of eight-hundred synthetic \ly{} spectra from our pool. We chose eight-hundred spectra because it provided a large enough sample such that the distribution for the value of $P$ at each $k$ approached a normal distribution. The process described in Section~\ref{sec:flux-powspec} was then applied to this bootstrapped sample in order to calculate one instance of the power spectrum.

\begin{figure*}
	\centering
    \includegraphics[width=\textwidth]{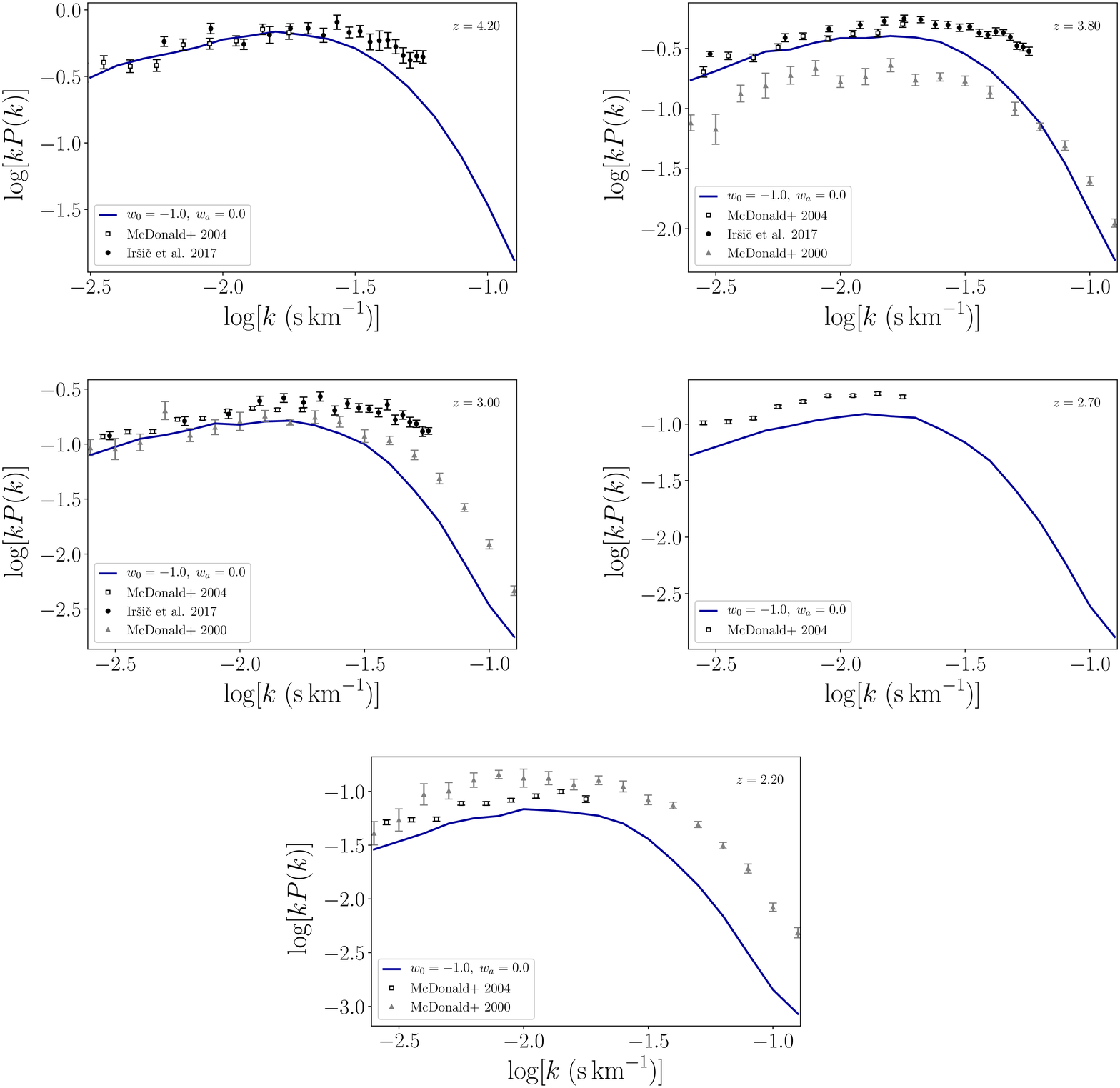}
    \caption{The power spectra from our L-40-1024 model compared to the observations of QSO absorbers at various redshifts of \citet{Irsic17,McDonald06,McDonald00}. There is an under-prediction of the flux power at the smallest scales. See text.}
    \label{fig:ps-with-obs}
\end{figure*}

    The power spectra from our L-40-1024 simulation are shown along with observational data from \citet{Irsic17,McDonald06,McDonald00} in \Cref{fig:ps-with-obs}. \Cref{fig:ps-with-obs} shows that the shape of the simulated power spectra matches the observations quite well at all redshifts, but there appears to be a scaling issue that results in the simulated power spectra under-predicting the power at large $k$ (small scales) and low redshift ($z = 2.7$ and $z = 2.2$ in particular). This result was also found by \citet{Bertone06} in their exploration of the effects of galactic winds on the \ly{} and in \citet{Peeples10} in their exploration of the effects of thermal broadening and heating rates on the \ly{}. See Section~\ref{sec:discussion} for a detailed discussion.

\subsubsection{Comparing Power Spectra}
\label{sec:compare-ps}
	The goal of this project was to analyze possible signatures of time-dependent dark energy in the flux power spectrum of the \ly{}. To this end, we needed a statistical test that could quantify the differences between our calculated power spectra. We chose the k-sample Anderson-Darling (AD) test \citep{anderson1952,Scholz87} for this comparison. The AD statistic is based upon the distance between the $k$ distributions being compared \citep{Scholz87}. We adopted the AD test for several reasons: first, it is distribution free\footnote{Meaning that no underlying distribution needs to be specified}; and second, when compared with the Kolmogorov-Smirnov (KS) test, the AD test puts more emphasis on the tails of the distribution, whereas the KS test emphasizes differences between distributions near the center. Since dark energy is a large-scale phenomenon, we expect most of the differences between power spectra, if they exist, to occur on the largest scales (smallest $k$), rather than in the central parts of the power spectrum; third, due to its increased sensitivity and ability to always be applied, the AD test has recently been recommended over the KS test in astronomy \citep[e.g.,][]{Babu06}.

\begin{deluxetable*}{lllllllll}
\tabcolsep=3pt
\tablecolumns{9}
\tablewidth{0pt}
\tablecaption{Anderson-Darling statistic for simulated power spectra as compared to the simulated cosmological constant power spectra \label{table:ad_table}}
\tabletypesize{\scriptsize}
\tablehead{\colhead{$z$} & \colhead{AD\tablenotemark{a}} & \colhead{C.V.\tablenotemark{a}} & \colhead{AD\tablenotemark{b}} & \colhead{C.V.\tablenotemark{b}} & \colhead{AD\tablenotemark{c}} & \colhead{C.V.\tablenotemark{c}} & \colhead{AD\tablenotemark{d}} & \colhead{C.V.\tablenotemark{d}}}
\startdata
    4.20 &    -1.23 &     1.96 &    -0.97 &     1.96 &    -1.18 &     1.96 &    -1.23 &     1.96 \\
    3.80 &    -1.23 &     1.96 &    -0.95 &     1.96 &    -1.23 &     1.96 &    -1.23 &     1.96 \\
    3.00 &    -1.23 &     1.96 &    -0.70 &     1.96 &    -1.23 &     1.96 &    -1.23 &     1.96 \\
    2.70 &    -1.23 &     1.96 &    -0.97 &     1.96 &    -1.23 &     1.96 &    -1.23 &     1.96 \\
    2.20 &    -1.23 &     1.96 &    -0.84 &     1.96 &    -1.23 &     1.96 &    -1.23 &     1.96 \\
\enddata
\tablecomments{C.V. is the critical value of the AD statistic at the chosen significance level ($\alpha = 0.05$).}
\tablenotetext{a}{$w_0 = 0.0,\; w_a = -3.0$}
\tablenotetext{b}{$w_0 = -1.1,\; w_a = 1.3$}
\tablenotetext{c}{$w_0 = -2.0,\; w_a = 0.0$}
\tablenotetext{d}{$w_0 = -2.0,\; w_a = 2.0$}
\end{deluxetable*}

    When comparing power spectra, we chose the L-40-1024 simulation as our fiducial simulation to compare to our other simulations. This is because the cosmological constant is the \emph{de facto} dark energy model in modern cosmology.

    The results of our comparison are given in \Cref{table:ad_table}. We used {\tt scipy} in order to conduct this test. The significance level $\alpha$ of the test represents the probability of a Type I error (i.e., the probability of rejecting the null hypothesis of a statistical test given that the null hypothesis is true). The null hypothesis of the test is rejected if the value of the test statistic is larger than the value of the critical value for the given significance level. \Cref{table:ad_table} shows that our AD statistics are lower than the critical values for $\alpha = 0.05$ in every case, indicating that one cannot reject the null hypothesis that the power spectra were drawn from the same distribution. This indicates that the intrinsic cosmic variance of the \ly{} power spectrum is in excess of the effects of time-dependent dark energy.

\begin{figure*}
	\centering
    \includegraphics[width=\textwidth]{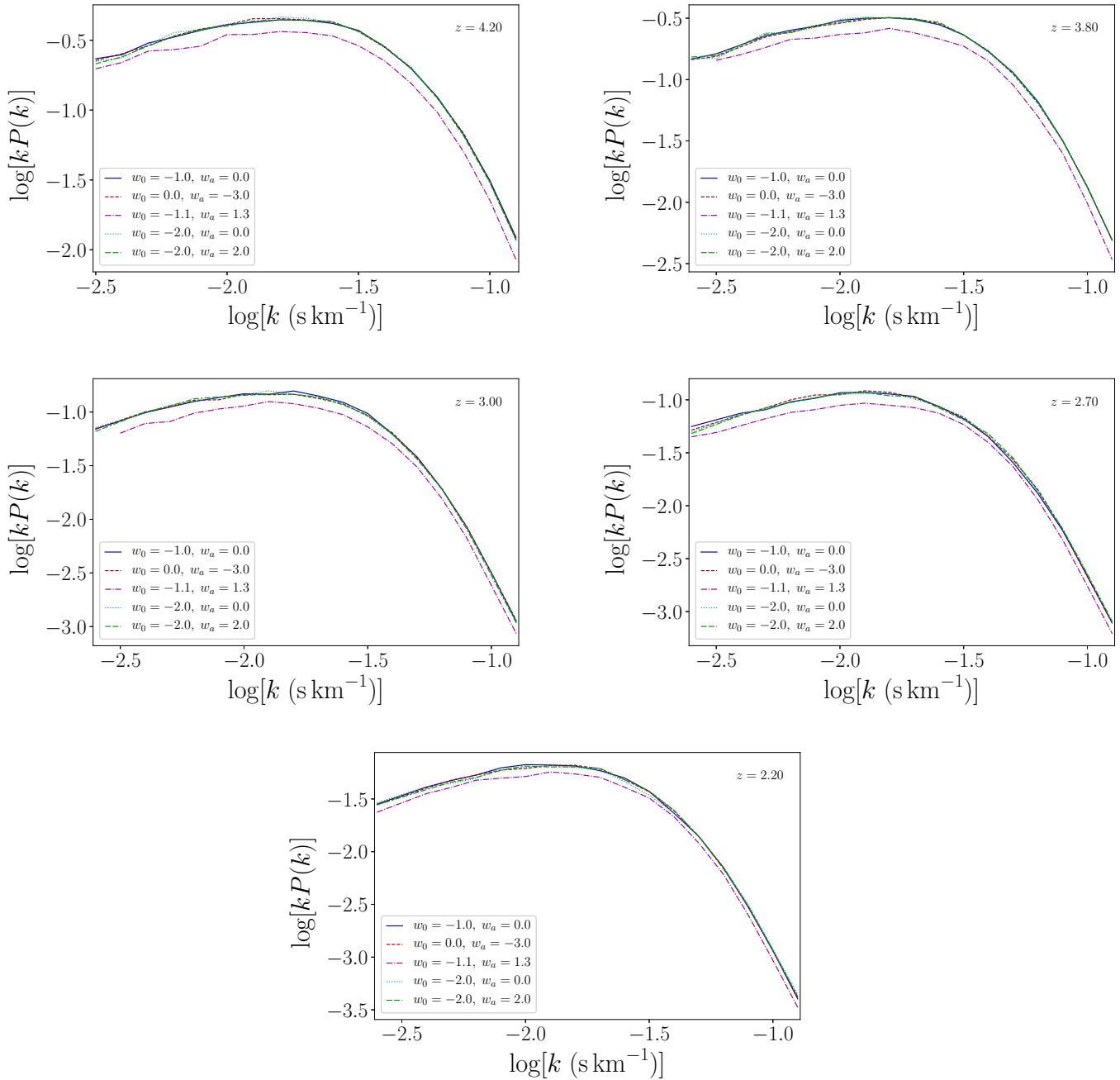}
    \caption{This figure shows the power spectra from each of our simulations. We see that the power spectra, save for that of DE2-40-1024, are very similar to one another. This reinforces the conclusion from our AD test.}
    \label{fig:all-ps}
\end{figure*}

\subsection{Large-Scale Region of the Power Spectra}
\label{sec:ls-ps-region}
\begin{figure*}
	\centering
    \includegraphics[width=\textwidth]{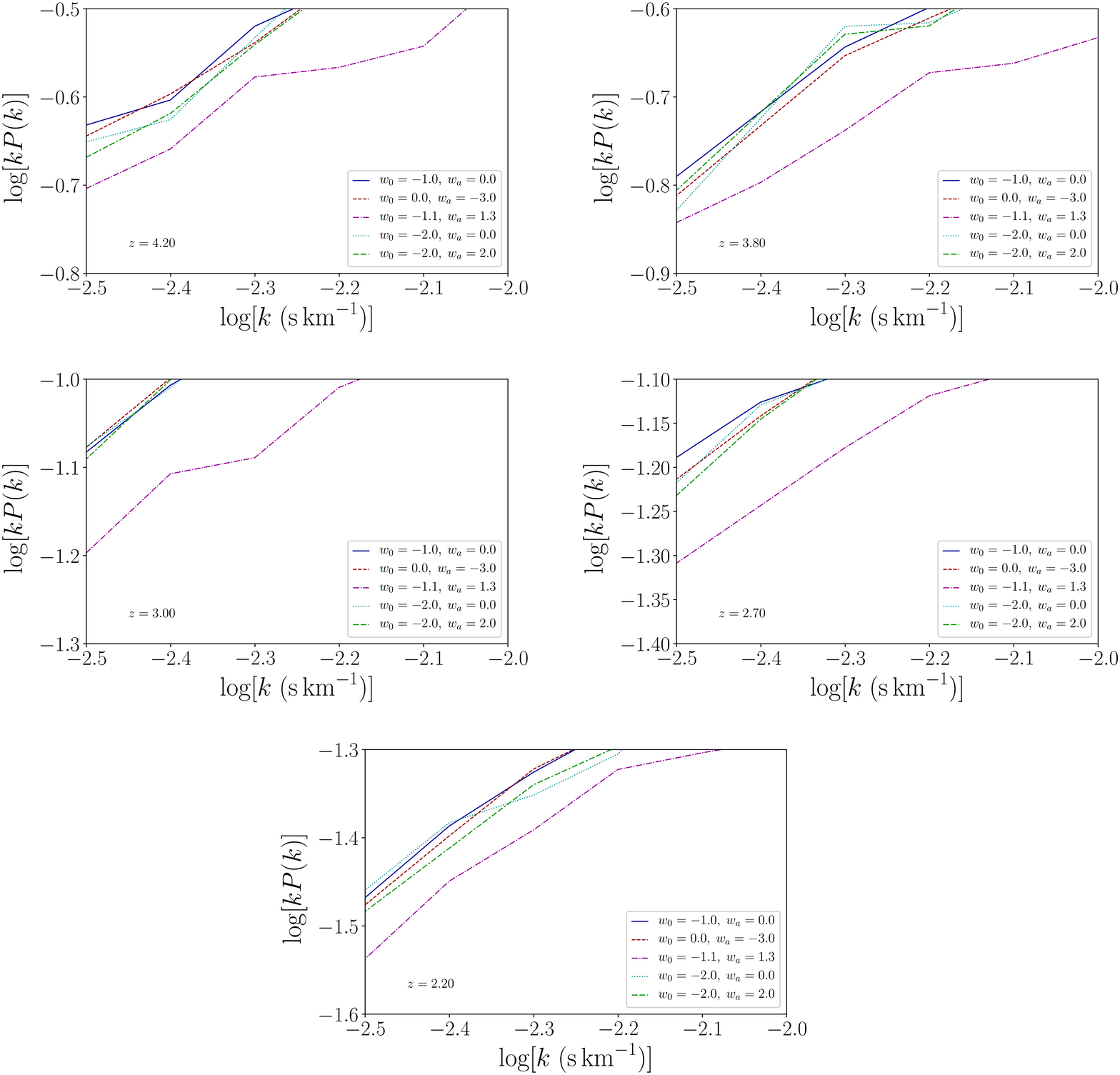}
    \caption{This figure is the same as \cref{fig:all-ps}, but zoomed in on the large-scale region of the power spectrum.}
    \label{fig:zoom-ps}
\end{figure*}

	Since dark energy is a large-scale phenomenon, we expect any signatures of time-dependent dark energy to be most prominent at small $k$. In order to investigate this, we zoomed in on each of the power spectra presented in \cref{fig:all-ps}, as seen in \cref{fig:zoom-ps}.
\Cref{fig:zoom-ps} shows that there are indeed small deviations in the power spectra at large scales.

\begin{figure*}
	\centering
    \includegraphics[width=\textwidth]{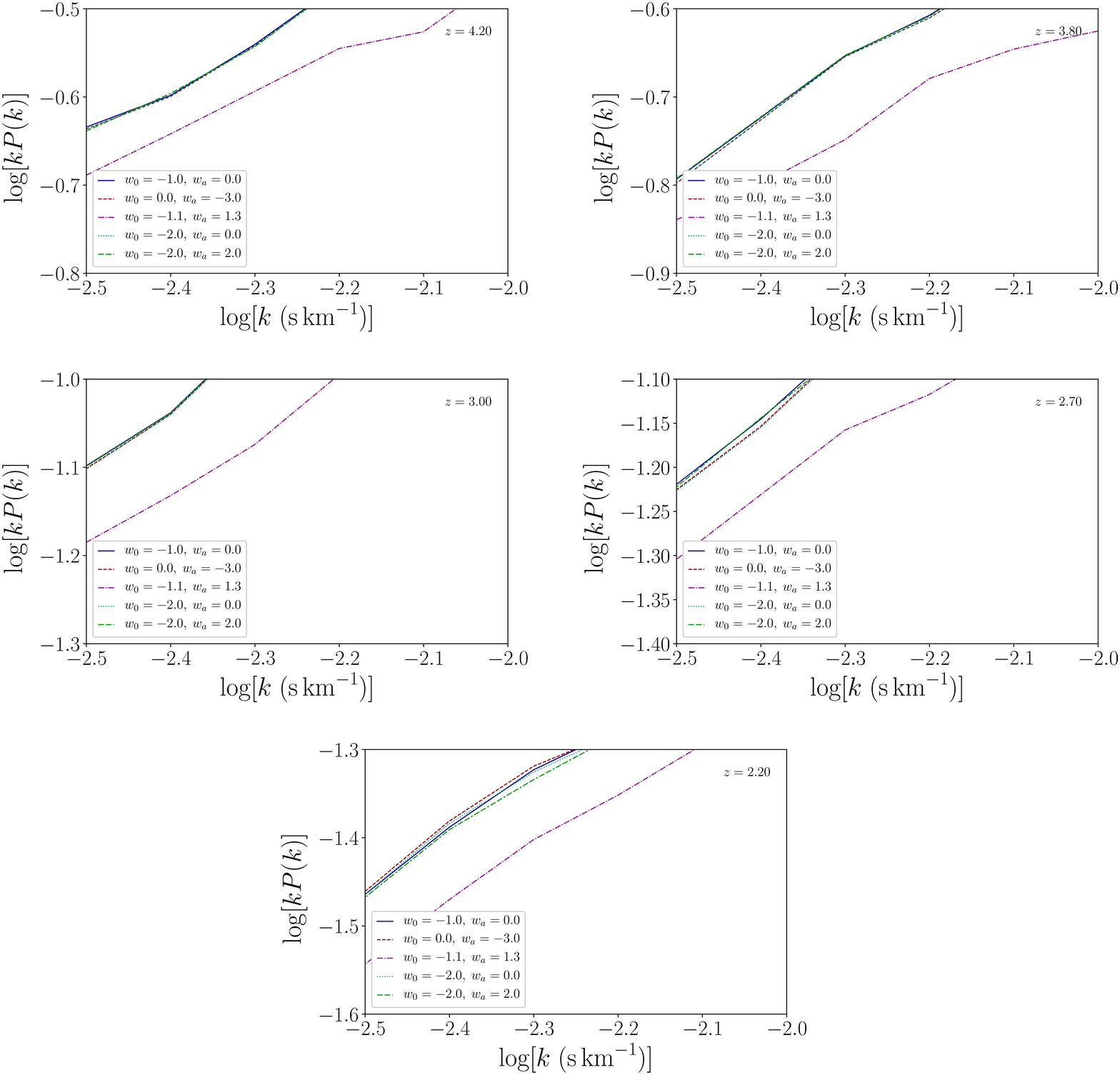}
    \caption{This figure shows the large-scale region of each synthetic power spectrum calculated from a pool of identical sight-lines, as opposed to randomly chosen sight-lines.}
    \label{fig:fixed-seed-zoom}
\end{figure*}

	However, our synthetic power spectra were calculated from a pool of synthetic \ly{} spectra, which contain randomly chosen sight-lines through our simulation volumes. In order to investigate whether or not these deviations are real and not merely statistical variations, we re-ran our spectral extraction code with a fixed random number generator seed. This guarantees that the pool of synthetic \ly{} spectra for each simulation all contain identical sight-lines. The synthetic power spectra arising from these pools of identical sight-lines are given in \cref{fig:fixed-seed-zoom}.

\Cref{fig:fixed-seed-zoom} shows that nearly all of the deviations between the flux power spectra vanish when using a pool of identical sight-lines. This indicates that the variations shown in \cref{fig:zoom-ps} are mostly due to variations in the pool of spectra. However, \cref{fig:fixed-seed-zoom} retains the discrepancy in power between the DE2-40-1024 and the other four models. This indicates that the power spectrum is sensitive to time-dependent dark energy, even if the differences between observationally viable models are small.

\subsection{Resolution Study}
\label{sec:res-study}
\begin{figure}
	\centering
    \includegraphics[width=\columnwidth]{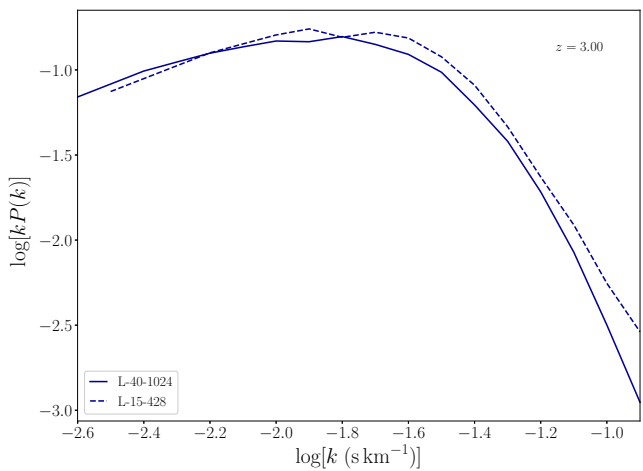}
    \caption{This figure shows the effects of using a smaller box size on the power spectrum. We have plotted the power spectrum for both the L-40-1024 (solid) and L-15-428 (dashed) simulations at $z = 3.00$. We see that at larger scales, the two power spectra are very similar, but as we move to smaller scales we see that there is an excess of power in the L-15-428 simulation.}
    \label{fig:res-box}
\end{figure}

	Here we investigate the effects of resolution on our results. To this end we ran four additional simulations, the parameters of which are given in \Cref{table:res-study-params}. These simulations cover several bases. The L-15-428 simulation has the same spatial resolution as our fiducial simulations, but with a smaller box size. The effects of this smaller box size are shown in \Cref{fig:res-box}. We see that at larger scales the two power spectra are nearly identical, but at smaller scales, the L-15-428 simulation has an excess of power compared to the L-40-1024 simulation.

\begin{figure}
	\centering
    \includegraphics[width=\columnwidth]{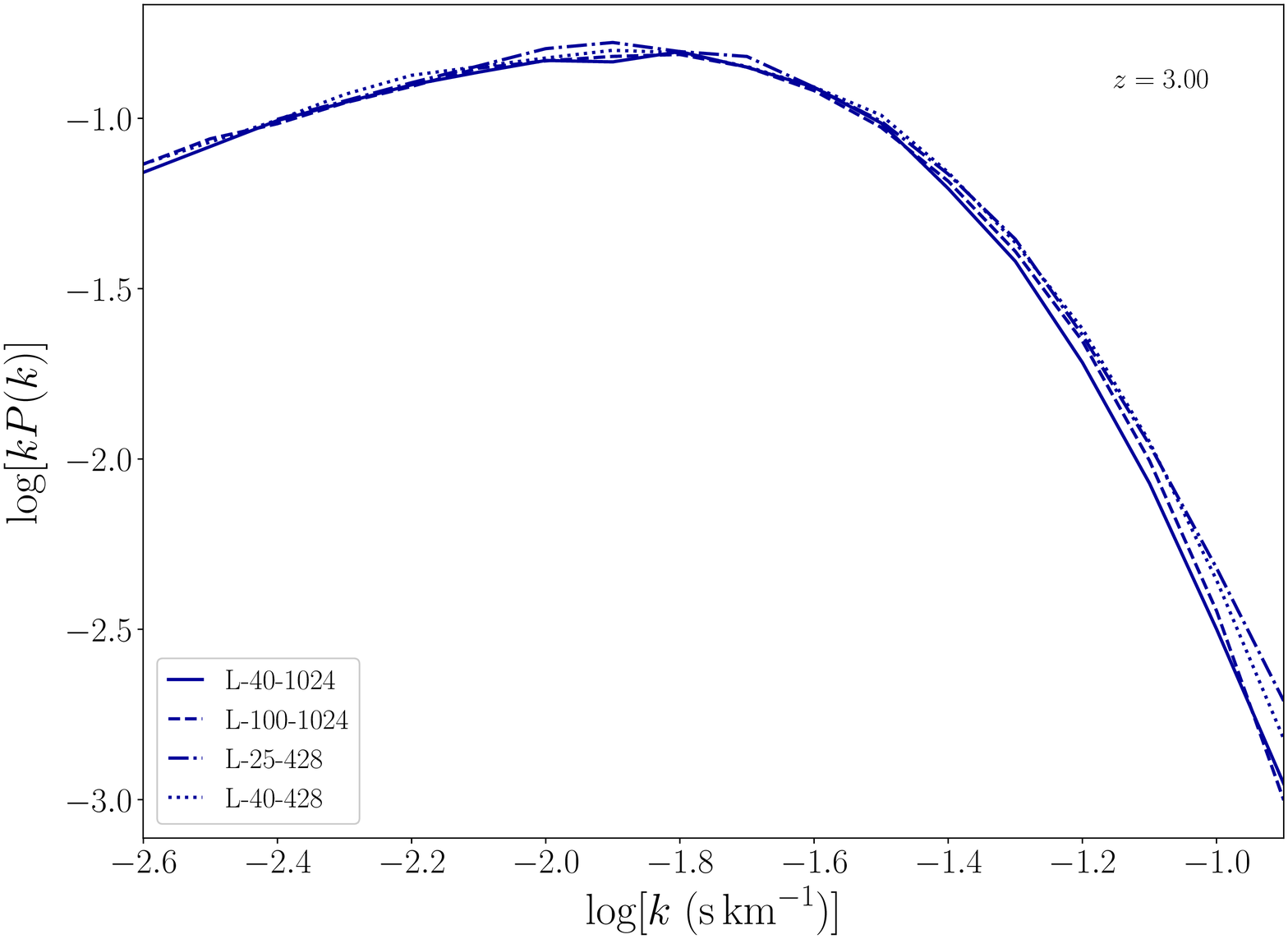}
    \caption{This figure shows the effects of varying the spatial resolution on the power spectrum. Here we have plotted our L-40-1024, L-100-1024, L-25-428, and L-40-428 models at $z = 3.0$. We see that at large scales, the models are similar to one another. The discrepancies arise at smaller scales. As expected, our L-100-1024 model, which has the largest box size, produces the least amount of power at small scales due to being unable to fully resolve the small-scale structure. The L-20-428 model produces the most power at small scales owing to its higher spatial resolution.}
    \label{fig:res-plot}
\end{figure}

    In addition to the L-15-428 simulation, we also have three other simulations, L-100-1024, L-25-428, and L-40-428, at varying spatial resolutions. These simulations are plotted alongside the fiducial L-40-1024 simulation in \Cref{fig:res-plot}. We see that at large scales each of the power spectra are nearly indistinguishable from one another. At smaller scales, however, there is a fairly substantial difference. The L-100-1024 simulation has under-predicted power whereas the L-25-428 simulation has over-predicted power as compared to the others. The L-40-428 and L-40-1024 simulations are nearly identical at all scales.

\begin{deluxetable*}{lllll}
    \tablewidth{0pt}
    \tabletypesize{\scriptsize}
    \tablecolumns{2}
    \tablecaption{Resolution Study Simulation Parameters\label{table:res-study-params}}
    \tablehead{\colhead{Simulation Name} & \colhead{Number of DM Particles} & \colhead{Softening Length ($\mathrm{kpc\,h}^{-1}$ comoving)} & \colhead{Box Size ($\mathrm{Mpc\,h}^{-1}$ comoving)} & \colhead{Particle Mass}}  
    \startdata
        L-100-1024 & $1024^3$ & $5.80$ & $100$ & $8.14 \times 10^7 M_\sun h^{-1}$ \\
        L-15-428 & $428^3$ & $1.40$ & $15$ & $3.76 \times 10^6 M_\sun h^{-1}$ \\
        L-25-428 & $428^3$ & $2.33$ & $25$ & $1.74 \times 10^7 M_\sun h^{-1}$ \\
        L-40-428 & $428^3$ & $3.73$ & $40$ & $7.13 \times 10^7 M_\sun h^{-1}$ \\
    \enddata
\end{deluxetable*}

We compared each of our resolution study simulations to our fiducial L-40-1024 simulation using the AD test, just as we did for our main production simulations. The results of this test are given in \cref{table:ad_table_res}. The results indicate that, although there are visual differences between the power spectra a various resolutions, the statistical evidence is not strong enough to discriminate between the power spectra at the $\alpha = 0.05$ significance level, indicating that the effects of resolution do not affect our conclusions.

\begin{deluxetable*}{llllllll}
\tabcolsep=3pt
\tablecolumns{8}
\tablewidth{0pt}
\tablecaption{Anderson-Darling statistic for resolution study power spectra as compared to the fiducial simulated power spectra at $z = 3.0$ \label{table:ad_table_res}}
\tabletypesize{\scriptsize}
\tablehead{\colhead{AD\tablenotemark{a}} & \colhead{C.V.\tablenotemark{a}} & \colhead{AD\tablenotemark{b}} & \colhead{C.V.\tablenotemark{b}} & \colhead{AD\tablenotemark{c}} & \colhead{C.V.\tablenotemark{c}} & \colhead{AD\tablenotemark{d}} & \colhead{C.V.\tablenotemark{d}}}
\startdata
   -1.21 &     1.96 &    -0.95 &     1.96 &    -1.17 &     1.96 &    -1.23 &     1.96 \\
\enddata
\tablecomments{C.V. is the critical value of the AD statistic at the chosen significance level ($\alpha = 0.05$).}
\tablenotetext{a}{L-100-1024}
\tablenotetext{b}{L-15-428}
\tablenotetext{c}{L-25-428}
\tablenotetext{d}{L-40-428}
\end{deluxetable*}

\section{Discussion}
\label{sec:discussion}
	This work has built upon the work performed in \citetalias{Viel03} in order to explore whether or not time-dependent dark energy leaves an observationally detectable signature in the flux power spectrum of the \ly{}. To this end, we extracted synthetic \ly{} spectra from high-resolution N-body simulations and used these to calculate the flux power spectrum. We used five different dark energy models, including the cosmological constant and four dynamical, parameterized dark energy models. These models were chosen from the $(w_0, w_a)$ posterior distributions as determined by Planck. In particular, of the four dynamical dark energy models we employed, three of them were chosen to lie at the fringes of this posterior, while the fourth was deliberately chosen to lie outside of the bounds determined by Planck to serve as an extreme example from which one might see a discernible effect on the power spectrum.

    Bootstrapping from our pool of synthetic \ly{} spectra, we calculated a power spectrum for each dark energy model at each redshift. We then utilized the k-sample AD test in order to compare our power spectra arising from simulations with dynamical dark energy to the power spectrum from our simulation employing a cosmological constant. The AD test can always be applied and emphasizes the tails of the distributions where we expect the effects of dark energy to be most prominent (at small $k$). The results of these tests show that there is no statistically significant evidence to discriminate between the power spectra at the $\alpha = 0.05$ significance level.

    This conclusion differs from the findings of \citetalias{Viel03}, who determined that the dark energy models they employed could, in principle, be discriminated against using the optical depth power spectrum. It should be noted, however, that \citetalias{Viel03} used dark energy models that all had a constant EOS different from $w(a) = -1$, whereas we have used dynamical models. Furthermore, the simulations employed by \citetalias{Viel03} were semi-analytic in nature whereas ours were N-body. In particular, we could determine that the statistical cosmic variance of the \ly{} power spectrum exceeds that of the signature of time-dependent dark energy. These two factors would seem to be consequential enough to account for our differences in conclusions.

    There are several points that affect our analysis. First is the fact that, given our high resolution requirements, each of our simulations was strictly N-body out of consideration for the total run time. This necessitated that we calculate the hydrodynamic quantities (such as density and temperature) required for calculating a synthetic spectrum in post-processing. In particular, our temperature prescription is an empirically determined power law, which does not encapsulate all of the necessary physical processes going on in the IGM, such as self-shielding, radiative cooling, shocks due to stellar winds and supernovae, metal feedback, etc. These omissions may affect our temperature-density relation, which could have an impact on the small-scale power where the discrepancy between the calculated and observed power spectra was the greatest. In particular, \citet{Bertone06} showed that the effects of galactic winds on the small-scale power are non-trivial. However, since we are concerned with larger scales that are not as affected by baryonic physics and galactic winds, omission of these effects in our simulations should not affect our conclusions (see, however, \cite{Bolton17} for the effects of these processes on the low-redshift flux power spectrum).

    Further, \citet{Peeples10} investigated the effect of various heating rates on the \ly{} and found that the thermal state of the IGM has a non-negligible effect on the \ly{}. In particular, their largest effect came from the inclusion of thermal broadening, which we also include. We note that these issues predominantly affect small-scale power, which we under predict. However, their effect on larger scales is not as significant, and so, while they could alter our power spectra, we do not expect the changes to be significant enough to affect our statistical conclusions.

    A second aspect of our simulations that needs to be addressed is the box size. Since dark energy is a large-scale phenomenon, ideally one would like to have as large a box as possible in order to explore smaller $k$ modes. However, while larger scales would be probed by a larger simulation volume, the \ly{} is predominantly due to absorption near clusters and inside filaments. This makes the effects of Hubble broadening on the lines and the correlation between absorbers separated by voids a secondary effect, making it difficult to detect.

    These points are highlighted in our resolution study. In every case, the effects of resolution on the power spectrum at large scales were negligible. While running in a larger box did probe larger scales, these scales are larger than those probed in the three observational datasets that we used in \Cref{fig:ps-with-obs}. Our study showed that the effects of resolution are most prominent on small-scale power, with lower resolution simulations under-predicting the observed small-scale power. This is in keeping with the findings of other authors \citep[e.g.,][]{McDonald03}.

    Interestingly, both of our simulations that were run in volumes of the same size (L-40-1024 and L-40-428) had very similar power spectra at all scales. This suggests that it is the box size, as opposed to the particle number, that has the largest resolution effect on the power spectrum.

    As a caveat to our dark energy model selection, we must be aware that the distribution we were sampling from for our $w_0$ and $w_a$ values is a posterior distribution, wherein all of the other relevant cosmological parameters have been marginalized. However, given the computational complexity of our simulations, such a Bayesian approach utilizing Markov-Chain Monte-Carlo (MCMC) methods in order to perform a similar marginalization was not feasible.

    Ideally, one would like to run a large grid of simulations from which MCMC analysis could be used to perform the same marginalization over the cosmological parameters, as was done in \citet{Planck15_DE}. While such a Bayesian approach would allow us to find the maximally likely $(w_0, w_a)$ contours when comparing simulation to observational data, our goal was to explore whether or not signatures from dark energy manifested themselves in the power spectrum  at a large enough level to be distinguished from the  cosmic variance as opposed to trying to find the best-fit dark energy model. This makes an MCMC approach of limited usefulness with regards to investigating the effects of time-dependent dark energy on the \ly{}.

    Finally, we note that the fact remains that the effects of dark energy on the \ly{} are sub-dominant to those of baryonic physics in the IGM. This implies that, whether performing a full marginalization over the cosmological parameters or not, the power spectra arising from varying time-dependent dark energy models will be extremely similar to one another. Our analysis shows that such models cannot be statistically discriminated against due to cosmic variance even at the level of idealized simulations where we are in possession of full knowledge of the physical situation. This means that any attempt to search for a dark energy signal in the observed flux power spectrum of the \ly{}, where the uncertainties are far larger, is, and likely will remain, challenging.

\acknowledgments
The authors would like to thank Serena Bertone for both useful discussions and generous help with development of the software used in this work. We would also like to thank the support team at the Center for Research Computing at the University of Notre Dame for their tireless help in getting the simulations performed for this work up and running. This work was supported by the U.S. Department of Energy under grant DE-FG02-95-ER40934.

\bibliography{references}
\end{document}